\UseRawInputEncoding

\documentclass[manuscript]{aastex631}
\usepackage{multirow}
\usepackage{ulem}

\shorttitle{Imaging preflare pulsations}
\shortauthors{Lv et al.}

\graphicspath{{./}{figures/}}

\newcommand\Rone{\uppercase\expandafter{\romannumeral1}}
\newcommand\Rtwo{\uppercase\expandafter{\romannumeral2}}
\newcommand\Rthree{\uppercase\expandafter{\romannumeral3}}

\begin{document}

\title{Imaging Preflare Broadband Pulsations in the Decimetric-metric Wavelengths}

\correspondingauthor{Yao Chen}
\email{yaochen@sdu.edu.cn}

\author{Maoshui Lv}
\affiliation{Institute of Space Sciences, Shandong University, Weihai, Shandong, 264209, People's Republic of China}

\author{Baolin Tan}
\affiliation{CAS Key Laboratory of Solar Activity, National Astronomical Observatories, Chinese Academy of Sciences, Beijing, 100101, People's Republic of China}
\affiliation{School of Astronomy and Space Science, University of Chinese Academy of Sciences, Beijing, 100049, People's Republic of China}

\author{Ruisheng Zheng}
\affiliation{Institute of Space Sciences, Shandong University, Weihai, Shandong, 264209, People's Republic of China}

\author{Zhao Wu}
\affiliation{Institute of Space Sciences, Shandong University, Weihai, Shandong, 264209, People's Republic of China}

\author{Bing Wang}
\affiliation{Institute of Space Sciences, Shandong University, Weihai, Shandong, 264209, People's Republic of China}

\author{Xiangliang Kong}
\affiliation{Institute of Space Sciences, Shandong University, Weihai, Shandong, 264209, People's Republic of China}

\author{Yao Chen}
\affiliation{Institute of Space Sciences, Shandong University, Weihai, Shandong, 264209, People's Republic of China}
\affiliation{Institute of Frontier and Interdisciplinary Science, Shandong University, Qingdao, Shandong, 266237, People's Republic of China}




\begin{abstract}

Preflare activities contain critical information about the pre-cursors and causes of solar eruptions. Here we investigate the characteristics and origin of a group of broadband pulsations (BBPs) in the decimetric-metric wavelengths, taking place during the preflare stage of the M7.1 flare dated on 2011 September 24. The event was recorded by multiple solar instruments including the Nan\c{c}ay Radioheliograh that measure the properties of the radio source. The BBPs start $\sim$24 min before the flare onset, extending from $<$ 360 to above 800 MHz with no discernible spectral drift. The BBPs consist of two stages, during the first stage the main source remains stationary, during the second stage it moves outward along with a steepening extreme-ultraviolet (EUV) wave driven by the eruption of a high-temperature structure. In both stages, we observe frequent EUV brightenings and jets originating from the flare region. During the second stage, the BBPs become denser in number and stronger in general, with the level of the polarization increasing gradually from $<$ 20{\%} to $>$ 60{\%} in the right-handed sense. These observations indicate the steepening EUV wave is important to the BBPs during the second stage, while the preflare reconnections causing the jets and EUV brightenings are important in both stages. This is the first time such a strong association of an EUV wave with BBPs is reported. We suggest a reconnection plus shock-sweeping-across-loop scenario for the cause of the BBPs.
%

\end{abstract}

\keywords{Solar coronal mass ejections (310), Solar activity (1475), Solar corona (1483), Solar flares (1496), Solar radio emission (1522)}
\section{Introduction}
Solar radio pulsations represent (quasi-)periodic or irregular short fluctuations observed from the radio flux curves or the dynamic spectra, in almost all wavelength ranges
from metric to microwave. In the decimetric-metric wavelengths, pulsations often appear as fine
structures being superposed on the wideband continuum of the type-IV solar radio burst (\citealt{1961ApJ...133..243Y}; \citealt{1973SoPh...29..461K}; \citealt{1977A&A....57..285D}; \citealt{1981A&A....93..129T}; \citealt{2002A&A...388..363K}. See the review by \citealt{2007LNP...725..251N}). According to the bandwidths of pulsations, they can be classified as broadband pulsations (BBPs) and narrowband pulsations (NBPs).

At least two aspects of the mechanisms of the radio pulsations should be considered. One is the radiation mechanism, the other is the cause of the pulsations/modulations. For NBPs a coherent mechanism of plasma radiation may be involved while for BBPs the gyrosynchrotron mechanism may be important. In addition, the Type-III like coherent radiation mechanism excited by fast-moving beam-type energetic electrons within large magnetic loops, as well as the loss-cone maser instability driven by trapped energetic electrons (\citealt{2003A&A...410.1001A}; \citealt{2003A&A...410.1011Z}. See \citealt{2020ApJ...891L..25N, 2021PhPl...28d0701N} and \citealt{2021ApJ...909L...5L} for latest theorical studies) have also been proposed as likely radiation mechanisms for BBPs. On the other hand, the pulsations/modulations are often explained with the following scenarios \citep{1980IAUS...86..341K, 1987SoPh..111..113A}: (1) modulations by the magnetohydrodynamic (MHD) oscillations in terms of, e.g., the fast sausage mode of the magnetic loop structure; (2) intrinsic oscillations of the kinetic radiation process involving the nonlinear wave-wave and wave-particle coupling; (3) modulations by the transient acceleration of energetic electrons, by, e.g., intermittent or bursty magnetic reconnections.

Most studies have focused on BBPs taking place during the impulsive or decay phases of solar flares. During the preflare stage, the energy releases, if exist, are much weaker than that during the main phase of the flares, thus the resultant radio signatures can be easily missed. \citet{2015ApJ...799...30Z} reported four flares with preflare microwave activities which contain fine structures such as quasi-periodic pulsations (QPPs) and millisecond dots, using the high-sensitivity data recorded by the Ond¡¦rejov radio spectrograph in the frequency range of 0.8-2.0 GHz. This reveals novel signatures of preflare activities. In a study on flare precursors in the low solar atmosphere, \citet{2017NatAs...1E..85W} found two episodes of weak emissions existing 10-30 min before the impulsive phase of the flare. The accompanying microwave spectra were recorded by the new Expanded Owens Valley Solar Array (EOVSA) in the frequency range of 2-18 GHz. They show that these precursor microwave emissions can be well modelled as quasi-thermal, gyrosynchrotron emission sources, and the spectral fittings have been used to deduce the magnetic field strengths and their temporal variation. This provides important constraint on the location and characteristics of the energy release from the radio perspective. Another study identified preflare microwave QPPs with the 17 GHz data recorded by the Nobeyama Radioheliograph \citep{2020ApJ...893L..17L}, in which the QPPs appear within the flare source region according to the microwave imaging data. Their periods increase from $\sim$ 300 to 500s. This has been used to infer the property of electric currents of the pre-flare source region.

Here we report a rare decimetric-metric event of BBPs observed during the preflare stage of an M7.1 limb flare. Multi-wavelength data, including both EUV and radio imaging/spectral data, are available. The close-to-limb perspective assures minor projection effect. The event provides a good opportunity to investigate the origin of BBPs. The observational data and results are presented in Sections 2 and 3. Conclusions and discussion are presented in Section 4.

\section{Data and Event Overview}

The flare occurred on 2011 September 24, starting at 12:33 UT, peaking at 13:17 UT, and ending at 14:10 UT, according to the \emph{Geostationary Operational Environmental Satellite} (\emph{GOES}) flare list. It originated from the NOAA AR 11302 close to the northeastern limb of the solar disk. Figure 1(a) shows the 171 {\AA} image observed by the Atmospheric Imaging Assembly (AIA; \citealt{2012SoPh..275...17L}) on board the \emph{Solar Dynamics Observatory} (\emph{SDO}; \citealt{2012SoPh..275....3P}) at the flare peak time, from which overlying loops are clearly visible. Mainly the EUV data at 171 {\AA} (Fe IX, $\sim$0.6 MK), 193 {\AA} (Fe XII, $\sim$1.6 MK), and 94 {\AA} (Fe XVIII, $\sim$6 MK) were analyzed here. The EUV data have a pixel size of 0.6\arcsec\ and a cadence of 12s.
An accompanying halo CME was observed by the Large Angle and Spectrometric Coronagraph (LASCO; \citealt{1995SoPh..162..357B}) C2 onboard the \emph{Solar and Heliospheric Observatory} (\emph{SOHO}; \citealt{1995SoPh..162....1D}) first at 12:48 UT (see Figure 1(b)). According to the second-order polynomial fit of the height measurements by C2, the CME starts at $\sim$12:33 UT.

The radio bursts were recorded by the spectrographs of e-Callisto Bleien (175-870 MHz) with a time resolution of 0.25s (Figure 1(c)). The bursts took place in all stages of the flare, from the preflare and early-rising stages to the impulsive and decay stages. \citet{2018SoPh..293...58L} have analyzed the stationary type IV continuum burst (12:40-13:00 UT) during the early-rising stage. Here we focus on the radio bursts observed from $\sim$12:09-12:31 UT before the onset of the flare.

The Nan\c{c}ay Radioheliograh (NRH; \citealt{1997LNP...483..192K}) provides imaging data with polarization measurement at 10 frequencies from 150 to 445 MHz. The spatial resolution depends on frequency and time of observation, being $\sim$2\arcmin\ at 445 MHz and $\sim$6\arcmin\ at 150 MHz in summer and up to three times larger along the NS direction during winter. Mainly the NRH data at 4 frequencies (360, 408, 432, and 445 MHz) were analyzed here.

The magnetic-field data on the photosphere are from the Helioseismic and Magnetic Imager (HMI; \citealt{2012SoPh..275..207S}) on board \emph{SDO}, with a pixel scale of 0.6\arcsec\ and a cadence of 45s. The coronal magnetic field configuration is extrapolated with the Potential Field Source Surface (PFSS; \citealt{2003SoPh..212..165S}) model.

\section{Preflare BBPs and Coronal Activities}

\subsection{Spectral Characteristics and Source Properties of the BBPs}
From Figure 2(a), the preflare BBPs started from $\sim$12:09 and ended at $\sim$12:31 UT, being $\sim$24 min before the flare onset. They extend from $<$ 360 to above 800 MHz with no discernible spectral drift. Hundreds of decimetric-metric BBPs can be identified. They are rather intermittent from $\sim$12:09 to 12:17 UT, and become denser in number and stronger in intensity later. There exist several minutes of weak emission at $\sim$12:17 UT. This has been used to split the whole BBPs into two stages, with stage I for $\sim$12:09 to 12:17 UT and stage II for $\sim$12:17-12:31 UT. There exists some background continuum that also becomes stronger during stage II. In addition, the overall bandwidth of relatively strong emission becomes wider in the later stage. During most time of this stage, the bandwidth is $>$500 MHz that is close to the emission frequency. This is why we classify the emission as BBPs.

Figure 2(b) presents the temporal curves of the brightness temperature ($T_b$) at four NRH frequencies (360, 408, 432, and 445 MHz). In accordance with the spectral data, there exist plenty of local peaks of $T_b$. Five blue vertical and dashed lines are plotted to show the correspondence between the spectral data and temporal curves, so to confirm the absence of any observable spectral drift of each BBP. The $T_b$ curves at the four frequencies are similar to each other, with the values of $T_b$ being close for the three larger frequencies while they are much smaller at 360 MHz. This is consistent with the cutoff of the spectral data that is around 360 MHz. The values of $T_b$ range from $\sim$10$^7$ K to $4 \times 10^8$ K in stage I and ranges from $\sim$10$^7$ K to $>$10$^9$ K in stage II.

Figure 2(c) presents the temporal variations of the levels of polarization at the four NRH frequencies. The levels are close to each other, agreeing with the fact they belong to the same BBPs. Another significant observation is that the polarization remains at a weak level during stage I while increasing consistently up to strong levels ($\sim$60{\%}) at the end of stage II, which continues to increase to $\sim 100{\%}$ after 12:31 UT.

Figure 2(d) shows the results of the wavelet analysis of the trend-subtracted $T_b$ profile at the NRH 445 MHz, from which one can identify several periodic components. The $\sim$ 5 min signal lasts for only about 5 min, thus it is not convincing; the $\sim$ 3 min signal lasts for about 9 min ($\sim$12:14-12:23 UT). Their origin can be inferred from the three intermittent enhancements of the spectral data (Figure 2(a)). There also exist several periodic components at $\sim$1 min and $\sim$2 min. Comprehensive studies exist regarding the origin of QPPs (reviewed briefly in the introduction section), therefore they are not considered further here.

Figure 3 presents the NRH sources at the four frequencies (panels (a)-(d)). Panel (e) shows the corresponding time line overplotted on the spectra. From this figure and the accompanying movie, we observe a major source located around the equatorial part of the limb. The source persists at the four NRH frequencies during the pre-flare BBPs, indicating the existence of the background continuum radiation. In addition to this major source, another two sources located at the northern part of the images (out of the limb) appear intermittently during the later period of stage II . Linking the three sources, one can obtain a large-scale arcade-like structure.

The plus sign in each panel represents the centroid of the major source at 445 MHz, plotted for the convenience of comparison. The size of the source is about 150-200\arcsec\ according to the outmost contours which represent the 30{\%} level of the maximum $T_b$ (the 30{\%} contours for short), close to the corresponding spatial resolution of NRH. The source centroids of the four frequencies are close to each other. During stage I the sources remain basically stationary (being $\sim$ 40\arcsec\ above the solar limb) while during stage II the sources move outward systematically. This can be seen from Figure 4(a) which presents the source centroid locations with the 85{\%} contours at different times. The dashed line that connects the centroids of the three frequencies (see Figure 4(a)) has been used as the slit to obtain the distance-time ($d-t$) measurement of the radio sources with the NRH data. The obtained $d-t$ data at 445 MHz are plotted in Figure 4(b), the error bars are given by the dimension of the 85{\%} contours of the source. They are overplotted onto the $d-t$ map of the AIA data at 193 {\AA} along the same slit that will be analyzed later. From the radio data, the speed of the radio source during stage I is not significant as expected, while the average speed increases to $\sim$120 km s$^{-1}$ around the end of stage II (from 12:28 to 12:31 UT).

\subsection{Coronal Activities and their relation with the BBPs}

Figure 5 and the accompanying movie present the AIA images at 171, 193, and 94 {\AA}. The left two columns present the AIA images at the start of stage I and around the end of stage II, respectively, the rightmost column presents the difference images. The three passbands are selected to show (1) the system of the coronal loops overlying the active region and the sporatic jets and EUV brightenings within/around the flare source region with 171 {\AA}, (2) the rise and steepening of the EUV wave structure with 193 {\AA}, and (3) the hot eruptive structure that drives the EUV wave with 94 {\AA}.

All these three aspects of the event are self-evident according to the AIA data. We plot arrows in Figure 5 to indicate their occurrence at the specific moment. What is important here is their physical connection to the BBP sources.

The jets move outward from the flare source toward the main radio source. The EUV brightenings take place frequently during the BBPs. The two phenomena indicate the occurrence of intermittent magnetic reconnection during the pre-flare stage. This can explain the similar intermittent behaviour of BBPs if attributing the radio-emitting energetic electrons to the same reconnection process.

The hot structure starts to rise around 12:10 UT according to the 94 {\AA} data. As mentioned, in Figure 4(b) we have plotted the $d-t$ map with the difference data at 193 {\AA}. Figure 4(c) presents the $d-t$ map at 94 {\AA} along the same slit. The rise of the ejecta consists of the gradual and the impulsive stage. The turning point of the two stages is at $\sim$12:25 UT. The steepening of the EUV wave also takes place around this moment (see Figure4(b)). According to the $d-t$ maps and the AIA images, the EUV wave emerges at $\sim$12:16 UT, which is always ahead of and thus is driven by the hot eruptive structure. They have the similar gradual-impulsive two-stage dynamics, with the speed of the EUV wave (hot structure) along the slit being $\sim$34 km s$^{-1}$ ($\sim$36 km s$^{-1}$) during the first stage and being $\sim$121 km s$^{-1}$ ($\sim$151 km s$^{-1}$) on during the second stage. Later, the EUV wave accelerates to $\sim$473 km s$^{-1}$ (see Figure 4(b)), indicating its steepening into a shock structure.

In Figure 6 and the accompanying movie we present the BBP sources at six NRH frequencies with the running-difference images at 193 {\AA}. The major source remains stationary first, after the appearance of the EUV wave the source is located at the EUV wave front and moves together with the EUV wave. This agrees with what has been observed from the $d-t$ map in Figure 4(b). Again, the arcade-like structure linking the three sources correlates well with the large-scale EUV wave front (see the white circles in Figure 6(g)).

\section{Conclusions and Discussion}

Here we presented detailed observations of the BBPs in the decimetric-metric wavelengths. The purpose is to understand the process causing the pulsations using multi-wavelength data including the radio-imaging data at several frequencies from NRH and the EUV imaging data from AIA/SDO. Such complete data set were reported for the first time ever for preflare decimetric-metric BBPs. The BBPs start $\sim$24 min before the major eruption, during which frequent EUV brightenings occur and several jets eject from the flare region towards the main radio source. During the early stage (stage I) the BBP sources remain stationary while during the later stage (stage II) the sources move outward with a steepening/acclerating EUV wave that is driven by an eruptive hot structure. The major BBP source, together with the other two sources, are located at the EUV wave front. In addition, the BBPs get intensified overall, and the level of polarization increases gradually upon the emergence of the EUV wave. These observations strongly indicate that the EUV wave contributes to the acceleration of the BBP-emitting energetic electrons during the later stage, in addition to the low-lying reconnection process that may provide seed particles to the EUV wave.

The combined action of the EUV wave and magnetic reconnection in releasing BBPs is demonstrated here for the first time. To see how this happens, in Figure 7(a) we show the HMI magnetogram superposed by (1) the magnetic field lines extrapolated with the usual PFSS method, (2) the EUV profile delineated with the 193 {\AA} data at 12:31:55 UT (also see the white circles in Figure 6(g)), and (3) the 50$\%$ contour of the major source observed by NRH at 12:31:57 UT. Note that the PFSS result suffers from the potential-field assumption and the usage of synoptic magnetogram, thus the extrapolated field lines should be treated with caution. Nevertheless, the overall morphology of the large-scale loop system agrees with the bright loops according to the 171 {\AA} images (see Figure 5). The location of the major BBP source correlates with the top of the loop system and the front of the EUV wave.

These observations are in line with the schematic of Figure 7(b). We suggest that during stage I the BBPs are radiated by energetic electrons that are released by the magnetic reconnection process taking place within the flare source region as evidenced by jets and EUV brightenings, during stage II after the EUV wave appearance these energetic electrons are further processed by the steepening wave and the resultant BBP source are carried outward by the EUV wave. In the meantime, the EUV wave sweeps a series of loop tops during its outward propagation. According to \citet{2015ApJ...798...81K,2016ApJ...821...32K}, such EUV (or shock)-wave-sweeping-loop-top process favors the acceleration of energetic particles since the loop tops serve as an efficient trapping agency within which particles can be processed by the steepening wave/shock-like structure for multiple times. The intensified pulsations during stage II are a natural result of this further processing, and the gradual increase of polarization level can be caused by the gradual change of the overall magnetic field orientation within the loop tops that are swept by the EUV wave front.

We tried to derive some key parameters (e.g., magnetic field, plasma density) in the radio sources through fitting the flux density spectra assuming the gyrosynchrotron emission. Such fittings require one to prescribe a group of free parameters, including the density of the background and energetic electrons, the field strength, the viewing angle, the column depth, and the size, etc. This means the obtained results are not unique and have very large uncertainty. In addition, the available number/range of frequencies with imaging data is quite limited here. For a proper fitting to deduce the source conditions, more observational constraints, such as measurement of the flux density at more frequencies and independent measurement of the magnetic field or plasma density, are required.

\begin{acknowledgments}
This study is supported by the National Natural Science Foundation of China (11973031, 11790303, and 11873036). The authors acknowledge the team of NRH for making their data available to us. We thank the Institute for Data Science FHNW Brugg/Windisch, Switzerland for providing the e-Callisto data.
\end{acknowledgments}

\bibliography{References}{}

\begin{thebibliography}{}
\expandafter\ifx\csname natexlab\endcsname\relax\def\natexlab#1{#1}\fi
\providecommand{\url}[1]{\href{#1}{#1}}
\providecommand{\dodoi}[1]{doi:~\href{http://doi.org/#1}{\nolinkurl{#1}}}
\providecommand{\doeprint}[1]{\href{http://ascl.net/#1}{\nolinkurl{http://ascl.net/#1}}}
\providecommand{\doarXiv}[1]{\href{https://arxiv.org/abs/#1}{\nolinkurl{https://arxiv.org/abs/#1}}}

\bibitem[{{Aschwanden}(1987)}]{1987SoPh..111..113A}
{Aschwanden}, M.~J. 1987, \solphys, 111, 113, \dodoi{10.1007/BF00145445}

\bibitem[{{Aurass} {et~al.}(2003){Aurass}, {Klein}, {Zlotnik}, \&
  {Zaitsev}}]{2003A&A...410.1001A}
{Aurass}, H., {Klein}, K.~L., {Zlotnik}, E.~Y., \& {Zaitsev}, V.~V. 2003, \aap,
  410, 1001, \dodoi{10.1051/0004-6361:20031249}

\bibitem[{{Brueckner} {et~al.}(1995){Brueckner}, {Howard}, {Koomen},
  {Korendyke}, {Michels}, {Moses}, {Socker}, {Dere}, {Lamy}, {Llebaria},
  {Bout}, {Schwenn}, {Simnett}, {Bedford}, \& {Eyles}}]{1995SoPh..162..357B}
{Brueckner}, G.~E., {Howard}, R.~A., {Koomen}, M.~J., {et~al.} 1995, \solphys,
  162, 357, \dodoi{10.1007/BF00733434}

\bibitem[{{Domingo} {et~al.}(1995){Domingo}, {Fleck}, \&
  {Poland}}]{1995SoPh..162....1D}
{Domingo}, V., {Fleck}, B., \& {Poland}, A.~I. 1995, \solphys, 162, 1,
  \dodoi{10.1007/BF00733425}

\bibitem[{{Droege}(1977)}]{1977A&A....57..285D}
{Droege}, F. 1977, \aap, 57, 285

\bibitem[{{Kai} \& {Takayanagi}(1973)}]{1973SoPh...29..461K}
{Kai}, K., \& {Takayanagi}, A. 1973, \solphys, 29, 461,
  \dodoi{10.1007/BF00150826}

\bibitem[{{Kerdraon} \& {Delouis}(1997)}]{1997LNP...483..192K}
{Kerdraon}, A., \& {Delouis}, J.-M. 1997, in Coronal Physics from Radio and
  Space Observations, ed. G.~{Trottet}, Vol. 483, 192,
  \dodoi{10.1007/BFb0106458}

\bibitem[{{Khan} {et~al.}(2002){Khan}, {Vilmer}, {Saint-Hilaire}, \&
  {Benz}}]{2002A&A...388..363K}
{Khan}, J.~I., {Vilmer}, N., {Saint-Hilaire}, P., \& {Benz}, A.~O. 2002, \aap,
  388, 363, \dodoi{10.1051/0004-6361:20020385}

\bibitem[{{Kong} {et~al.}(2016){Kong}, {Chen}, {Guo}, {Feng}, {Du}, \&
  {Li}}]{2016ApJ...821...32K}
{Kong}, X., {Chen}, Y., {Guo}, F., {et~al.} 2016, \apj, 821, 32,
  \dodoi{10.3847/0004-637X/821/1/32}

\bibitem[{{Kong} {et~al.}(2015){Kong}, {Chen}, {Guo}, {Feng}, {Wang}, {Du}, \&
  {Li}}]{2015ApJ...798...81K}
---. 2015, \apj, 798, 81, \dodoi{10.1088/0004-637X/798/2/81}

\bibitem[{{Kuijpers}(1980)}]{1980IAUS...86..341K}
{Kuijpers}, J. 1980, in Radio Physics of the Sun, ed. M.~R. {Kundu} \& T.~E.
  {Gergely}, Vol.~86, 341--360, \dodoi{10.1017/S0074180900037098}

\bibitem[{{Lemen} {et~al.}(2012){Lemen}, {Title}, {Akin}, {Boerner}, {Chou},
  {Drake}, {Duncan}, {Edwards}, {Friedlaender}, {Heyman}, {Hurlburt}, {Katz},
  {Kushner}, {Levay}, {Lindgren}, {Mathur}, {McFeaters}, {Mitchell}, {Rehse},
  {Schrijver}, {Springer}, {Stern}, {Tarbell}, {Wuelser}, {Wolfson}, {Yanari},
  {Bookbinder}, {Cheimets}, {Caldwell}, {Deluca}, {Gates}, {Golub}, {Park},
  {Podgorski}, {Bush}, {Scherrer}, {Gummin}, {Smith}, {Auker}, {Jerram},
  {Pool}, {Soufli}, {Windt}, {Beardsley}, {Clapp}, {Lang}, \&
  {Waltham}}]{2012SoPh..275...17L}
{Lemen}, J.~R., {Title}, A.~M., {Akin}, D.~J., {et~al.} 2012, \solphys, 275,
  17, \dodoi{10.1007/s11207-011-9776-8}

\bibitem[{{Li} {et~al.}(2021){Li}, {Chen}, {Ni}, {Tan}, {Ning}, \&
  {Zhang}}]{2021ApJ...909L...5L}
{Li}, C., {Chen}, Y., {Ni}, S., {et~al.} 2021, \apjl, 909, L5,
  \dodoi{10.3847/2041-8213/abe708}

\bibitem[{{Li} {et~al.}(2020){Li}, {Li}, {Lu}, {Zhang}, {Ning}, \&
  {Anfinogentov}}]{2020ApJ...893L..17L}
{Li}, D., {Li}, Y., {Lu}, L., {et~al.} 2020, \apjl, 893, L17,
  \dodoi{10.3847/2041-8213/ab830c}

\bibitem[{{Liu} {et~al.}(2018){Liu}, {Chen}, {Cho}, {Feng}, {Vasanth}, {Koval},
  {Du}, {Wu}, \& {Li}}]{2018SoPh..293...58L}
{Liu}, H., {Chen}, Y., {Cho}, K., {et~al.} 2018, \solphys, 293, 58,
  \dodoi{10.1007/s11207-018-1280-y}

\bibitem[{{Ni} {et~al.}(2021){Ni}, {Chen}, {Li}, {Sun}, {Ning}, \&
  {Zhang}}]{2021PhPl...28d0701N}
{Ni}, S., {Chen}, Y., {Li}, C., {et~al.} 2021, Physics of Plasmas, 28, 040701,
  \dodoi{10.1063/5.0045546}

\bibitem[{{Ni} {et~al.}(2020){Ni}, {Chen}, {Li}, {Zhang}, {Ning}, {Kong},
  {Wang}, \& {Hosseinpour}}]{2020ApJ...891L..25N}
---. 2020, \apjl, 891, L25, \dodoi{10.3847/2041-8213/ab7750}

\bibitem[{{Nindos} \& {Aurass}(2007)}]{2007LNP...725..251N}
{Nindos}, A., \& {Aurass}, H. 2007, in Lecture Notes in Physics, Berlin
  Springer Verlag, ed. K.-L. {Klein} \& A.~L. {MacKinnon}, Vol. 725, 251

\bibitem[{{Pesnell} {et~al.}(2012){Pesnell}, {Thompson}, \&
  {Chamberlin}}]{2012SoPh..275....3P}
{Pesnell}, W.~D., {Thompson}, B.~J., \& {Chamberlin}, P.~C. 2012, \solphys,
  275, 3, \dodoi{10.1007/s11207-011-9841-3}

\bibitem[{{Scherrer} {et~al.}(2012){Scherrer}, {Schou}, {Bush}, {Kosovichev},
  {Bogart}, {Hoeksema}, {Liu}, {Duvall}, {Zhao}, {Title}, {Schrijver},
  {Tarbell}, \& {Tomczyk}}]{2012SoPh..275..207S}
{Scherrer}, P.~H., {Schou}, J., {Bush}, R.~I., {et~al.} 2012, \solphys, 275,
  207, \dodoi{10.1007/s11207-011-9834-2}

\bibitem[{{Schrijver} \& {De Rosa}(2003)}]{2003SoPh..212..165S}
{Schrijver}, C.~J., \& {De Rosa}, M.~L. 2003, \solphys, 212, 165,
  \dodoi{10.1023/A:1022908504100}

\bibitem[{{Trottet} {et~al.}(1981){Trottet}, {Kerdraon}, {Benz}, \&
  {Treumann}}]{1981A&A....93..129T}
{Trottet}, G., {Kerdraon}, A., {Benz}, A.~O., \& {Treumann}, R. 1981, \aap, 93,
  129

\bibitem[{{Wang} {et~al.}(2017){Wang}, {Liu}, {Ahn}, {Xu}, {Jing}, {Deng},
  {Huang}, {Liu}, {Kusano}, {Fleishman}, {Gary}, \&
  {Cao}}]{2017NatAs...1E..85W}
{Wang}, H., {Liu}, C., {Ahn}, K., {et~al.} 2017, Nature Astronomy, 1, 0085,
  \dodoi{10.1038/s41550-017-0085}

\bibitem[{{Young} {et~al.}(1961){Young}, {Spencer}, {Moreton}, \&
  {Roberts}}]{1961ApJ...133..243Y}
{Young}, C.~W., {Spencer}, C.~L., {Moreton}, G.~E., \& {Roberts}, J.~A. 1961,
  \apj, 133, 243, \dodoi{10.1086/147019}

\bibitem[{{Zhang} {et~al.}(2015){Zhang}, {Tan}, {Karlick{\'y}},
  {M{\'e}sz{\'a}rosov{\'a}}, {Huang}, {Tan}, \&
  {Sim{\~o}es}}]{2015ApJ...799...30Z}
{Zhang}, Y., {Tan}, B., {Karlick{\'y}}, M., {et~al.} 2015, \apj, 799, 30,
  \dodoi{10.1088/0004-637X/799/1/30}

\bibitem[{{Zlotnik} {et~al.}(2003){Zlotnik}, {Zaitsev}, {Aurass}, {Mann}, \&
  {Hofmann}}]{2003A&A...410.1011Z}
{Zlotnik}, E.~Y., {Zaitsev}, V.~V., {Aurass}, H., {Mann}, G., \& {Hofmann}, A.
  2003, \aap, 410, 1011, \dodoi{10.1051/0004-6361:20031250}

\end{thebibliography}
\bibliographystyle{aasjournal}

\begin{figure*}[h]
 \centering
  \includegraphics[width=16cm]{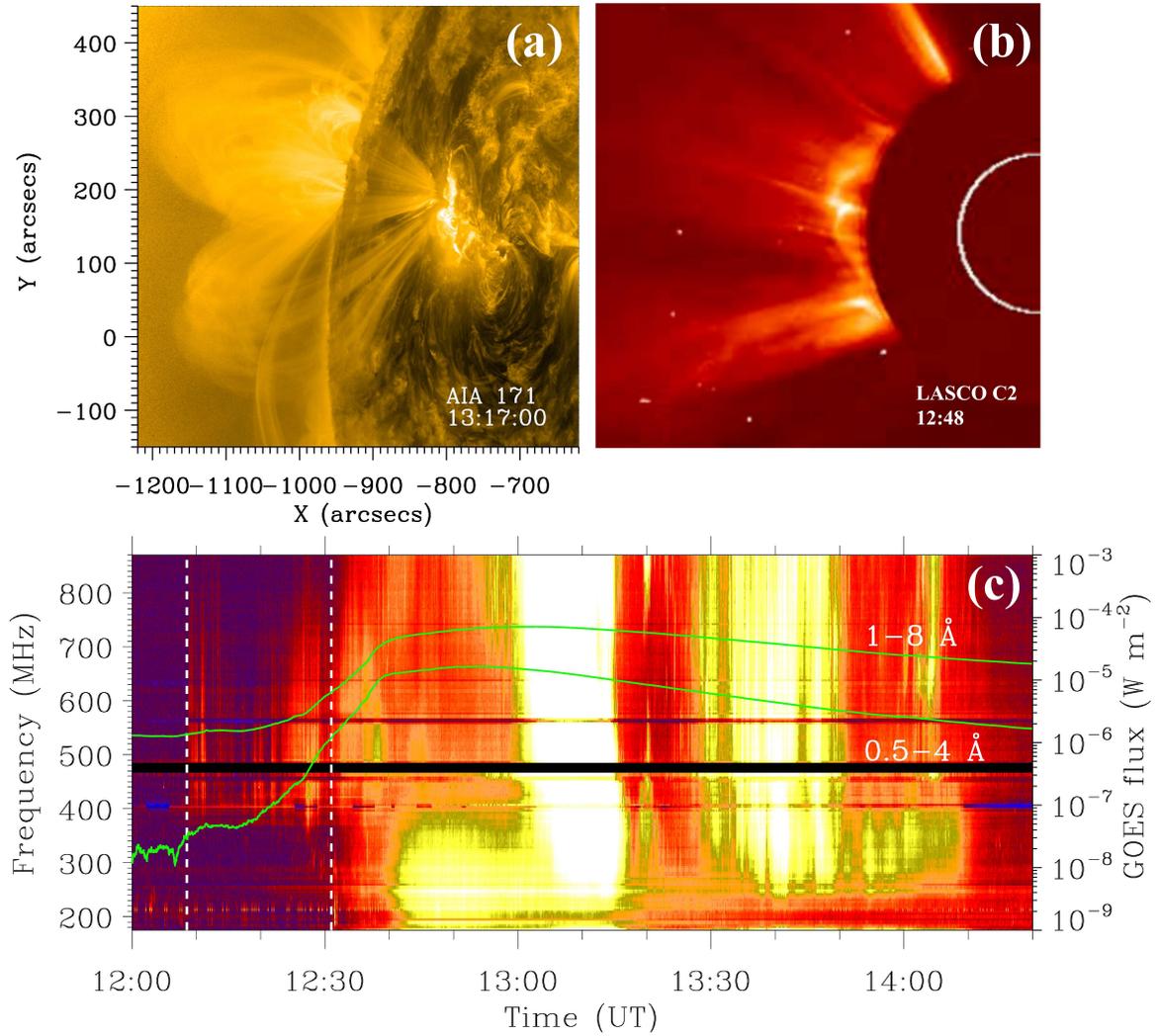}
  \caption{Overview of the Event: (a) the AIA 171 {\AA} data observed around the flare peak time, (b) the white light CME first observed by LASCO C2, (c) the dynamic spectra given by the e-Callisto Bleien observatory with the GOES 1-8 {\AA} and 0.5-4 {\AA} light curves. The two vertical dashed lines in panel (c) give the period with the preflare BBPs.}\label{fig1}
\end{figure*}

\begin{figure*}[h]
 \centering
  \includegraphics[width=16cm]{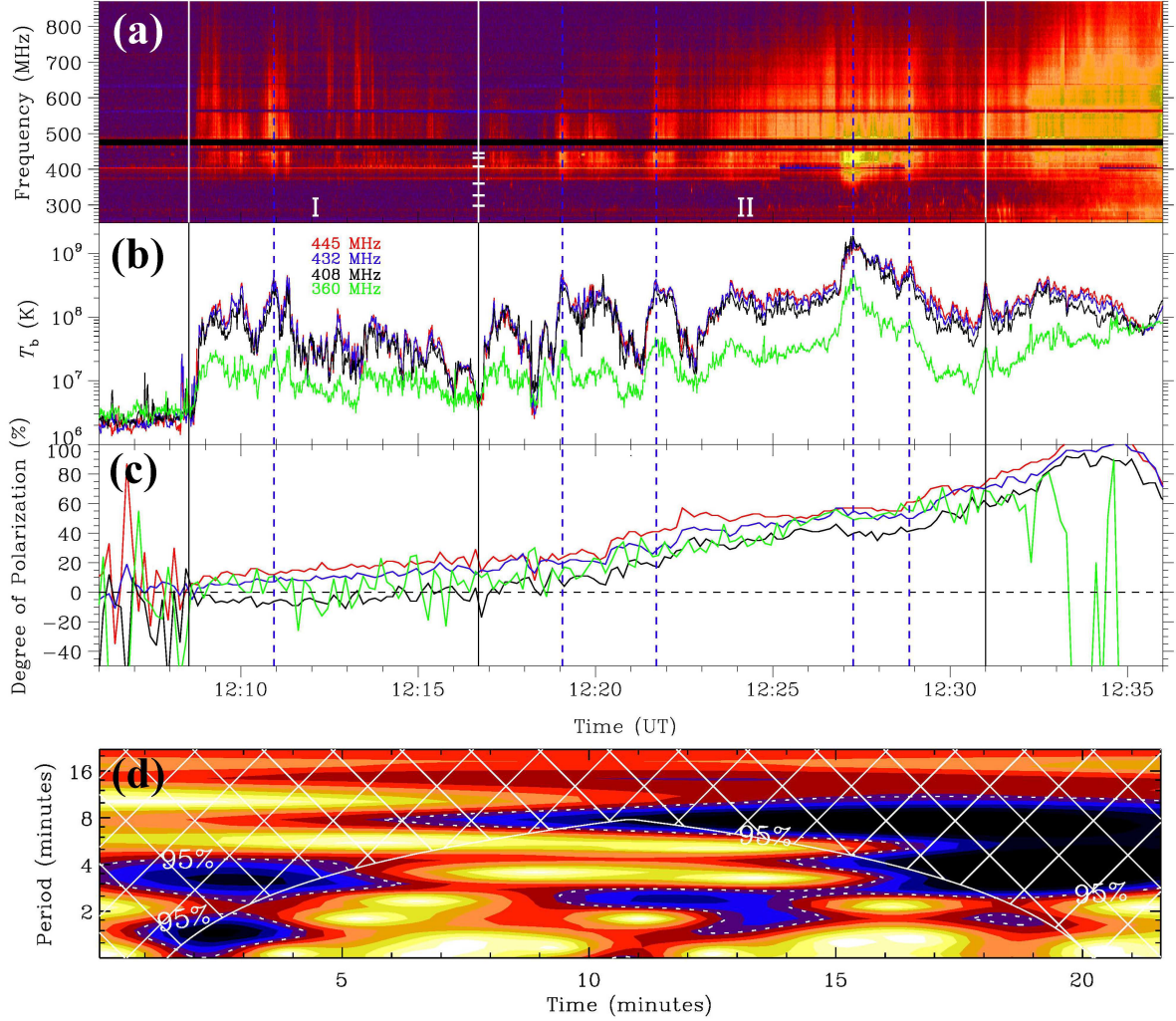}
  \caption{Overview of the preflare BBPs: (a) the dynamic spectra observed from 12:06 to 12:36 UT for the frequency range of [250, 870] MHz, (b) the temporal profiles of \emph{T}${\rm _b}$ at 1s cadence and (c) the degree of polarization with 10s cadence at 360, 408, 432, and 445 MHz, (d) the power spectrum given by the wavelet analysis using the \emph{T}${\rm _b}$ data at 445 MHz, starting from 12:09 UT. The \emph{T}${\rm _b}$ and the polarization level are obtained by averaging the corresponding data within the contour of 85\% of the maximum $T_b$. The two outmost vertical lines in panels (a)-(c) show the preflare period to be analyzed, the line with short horizontal lines (representative of the NRH frequencies) separate the BBPs into two stages (I and II), the blue vertical dashed lines indicate five local \emph{T}${\rm _b}$ peaks to show the absence of observable spectral drift. The dashed line in panel (d) gives the 95\% confidence level, and the cross-hatched region presents the cone of influence.}\label{fig2}
\end{figure*}

\begin{figure*}[h]
 \centering
  \includegraphics[width=16cm]{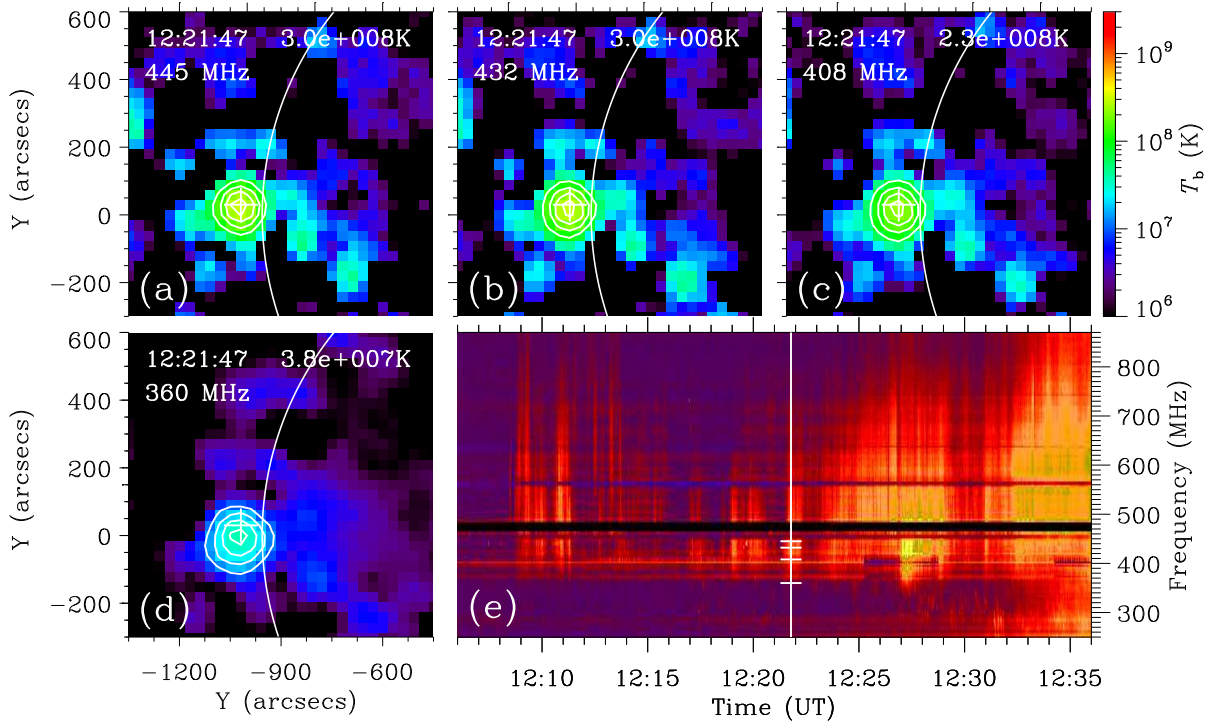}
  \caption{The radio sources observed by NRH at 360, 408, 432, and 445 MHz (a-d), and the corresponding dynamic spectra showing the time line (the vertical line) of the radio images (e). The contours in panels (a)-(d) are given by the 30, 50, 70, and 90\% levels of the corresponding maximum $T_b$ shown in each panel, and the plus sign represents the centroid of the source at 445 MHz for comparison. The short horizontal lines in panel (e) indicate the four NRH frequencies. An animation from 12:08 to 12:31 UT is available online, which shows the evolution of the BBP sources during the two stages. The real-time duration of the animation is 14 seconds.
}\label{fig3}
\end{figure*}

\begin{figure*}[h]
 \centering
  \includegraphics[width=18cm]{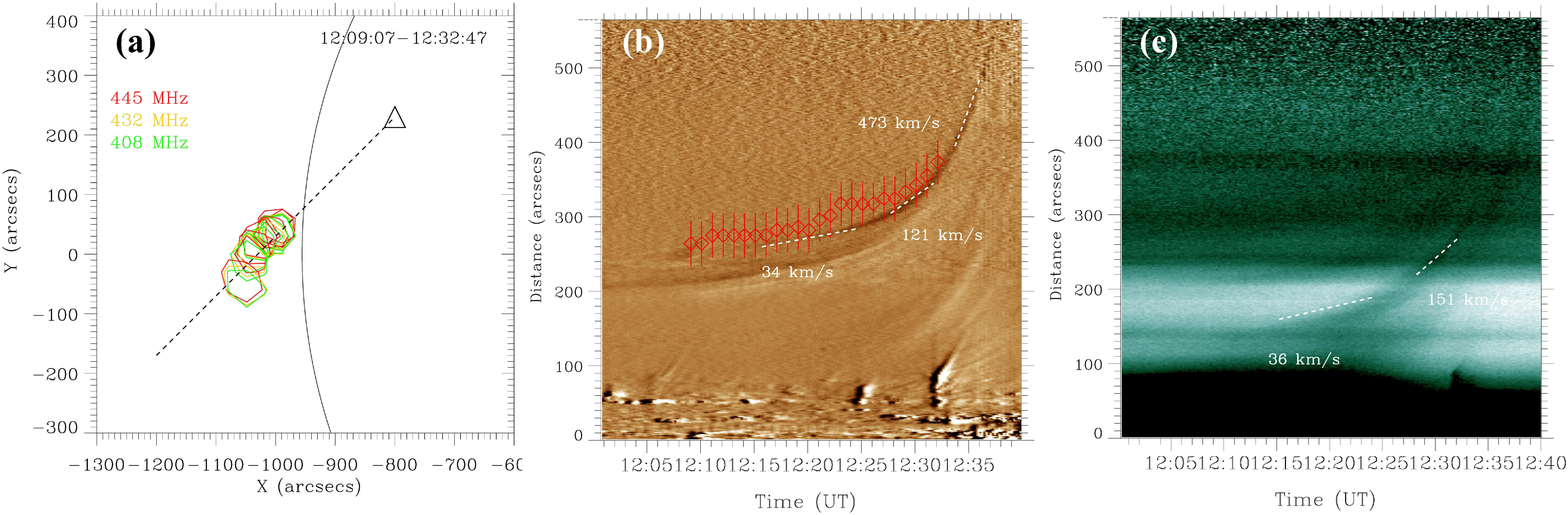}
  \caption{(a) The temporal variation of the 85\% contours of the sources at the three NRH frequencies to give the slit (or the moving direction) for further analysis, (b) the distance-time ($d-t$) maps with the 193 {\AA} running-difference data and (c) the direct 94 {\AA} data. The triangle in panel (a) represents the source of the flare. The diamonds with bars in panel (b) represent the $d-t$ data of the source centroid locations at 445 MHz, the bars are given by the dimension of the 85\% contours of the source. In panels (b)-(c), the $d-t$ data are given along the same slit presented in panel (a), and the dashed lines represent the linear fittings to calculate the speeds.}\label{fig4}
\end{figure*}

\begin{figure*}[h]
 \centering
  \includegraphics[width=16cm]{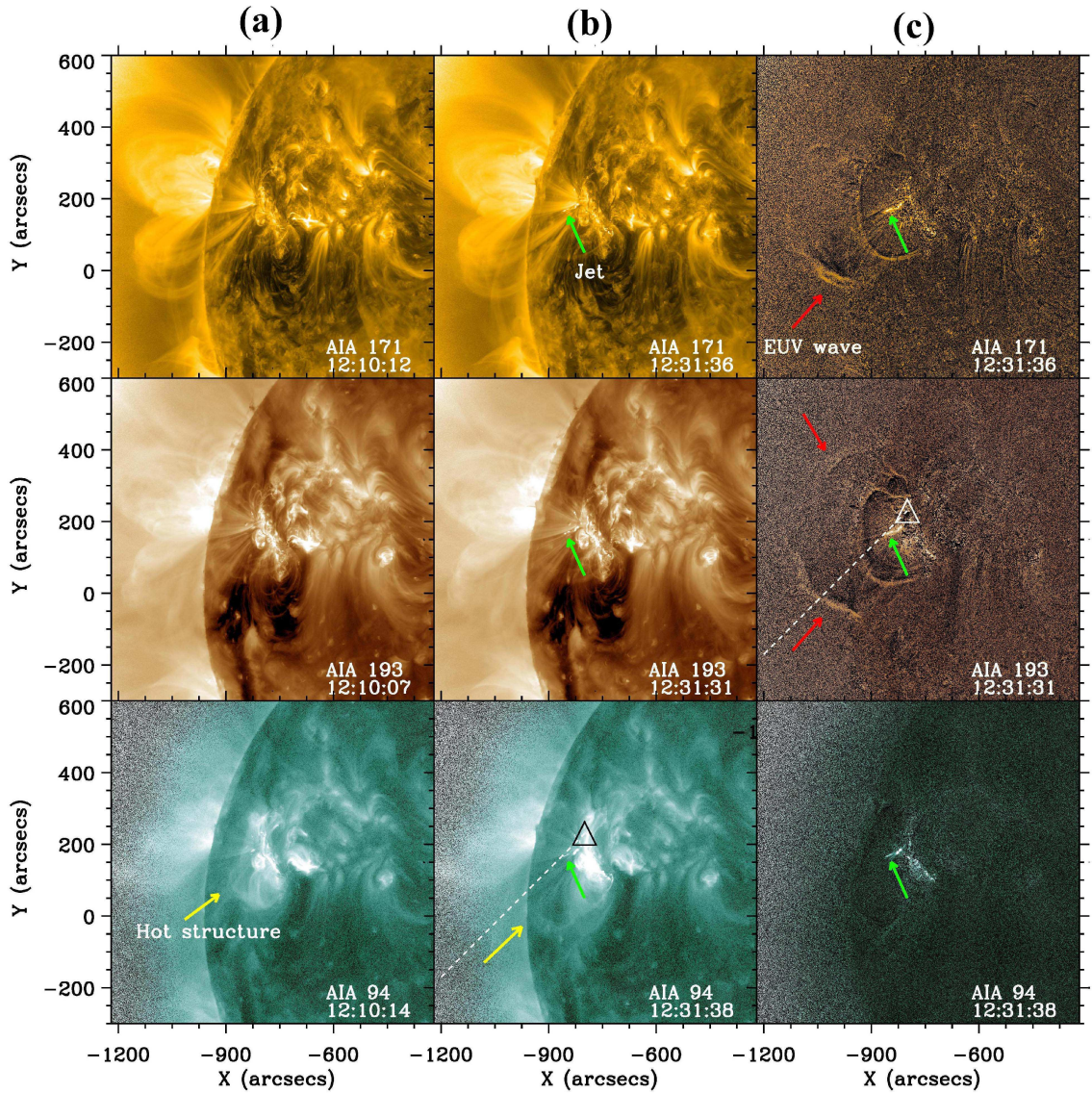}
  \caption{The AIA data of the event: (a) The 171, 193, and 94 {\AA} images observed at the start of stage I and (b) around the end of stage II, (c) the running-difference images observed around the end of stage II. The dashed line represents the slit taken from Figure 4(a) to give the $d-t$ maps of Figures 4(b)-(c), the triangle represents the flare source. The arrows point to the hot structure (yellow), the EUV wave (red), and the jets (orange). An animation from 12:05 to 12:35 UT is available online, which shows the brightenings and jets, and the evolutions of the hot structure and the EUV wave at 171, 193, and 94 {\AA} during the two stages. The real-time duration of the animation is 5 seconds.
}\label{fig5}
\end{figure*}

\begin{figure*}[h]
 \centering
  \includegraphics[width=16cm]{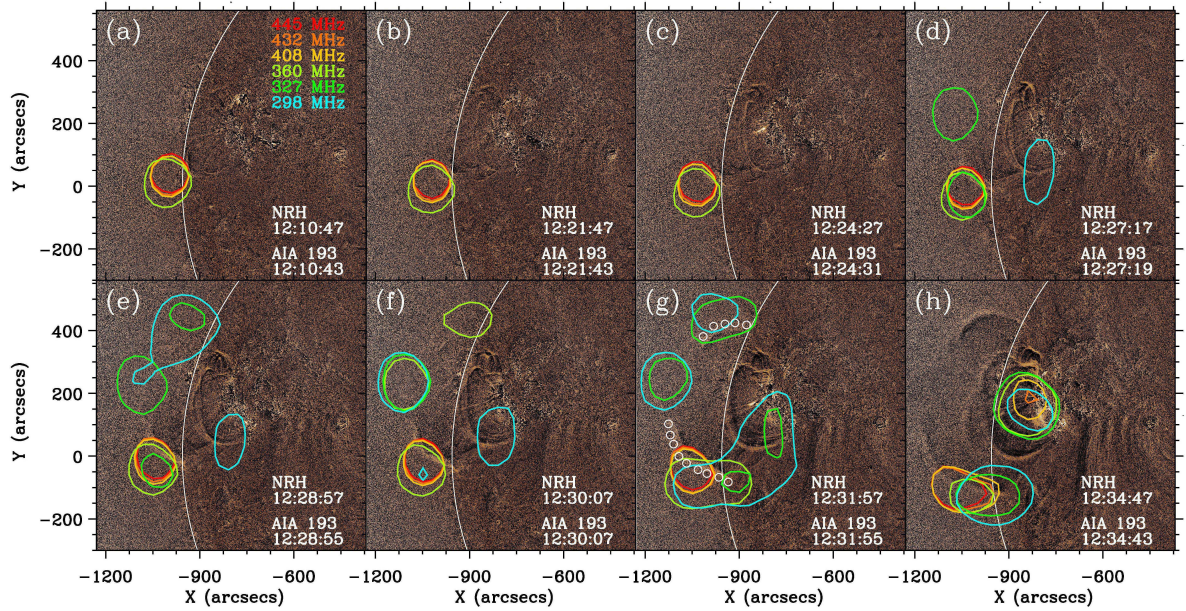}
  \caption{The combined AIA and NRH data at different times from 12:10 to 12: 34 UT. The running-difference images are at 193 {\AA}, and the NRH contours are at the 50\% levels for six NRH frequencies. The white circles in panel (g) delineate the front of the steepening EUV wave. An animation from 12:05 to 12:35 UT of the running difference images is available online. The animation shows the simultaneous evolution of the BBP sources and the EUV wave at 193 {\AA} during the two stages. The bottom portion of the animation includes the corresponding dynamic spectra showing the time line (the vertical line) of the radio images. This part of the animation is not shown in the Figure. The real-time duration of the animation is 5 seconds.
}\label{fig6}
\end{figure*}

 \begin{figure*}[h]
 \centering
  \includegraphics[width=16cm]{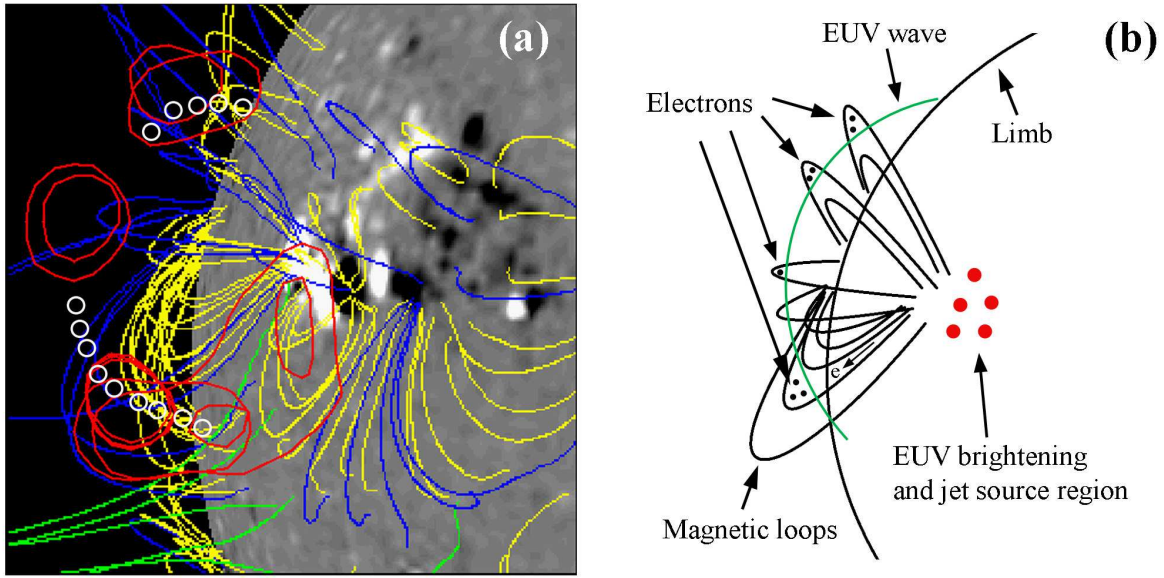}
 \caption{(a) The extrapolated PFSS field lines superposed onto the HMI magnetogram data, the contours are given by the 50\% level of the corresponding maximum $T_b$ at the six NRH frequencies, (b) the schematic to illustrate the physical origin of the BBPs. The white circles in panel (a) are taken from Figure 6(g) to show the front of the EUV wave.}\label{fig7}
\end{figure*}

\end{document}